\documentclass[11pt]{article}

\usepackage{amsmath,amssymb,amsfonts}
\DeclareMathOperator*{\argmax}{arg\,max}
\usepackage{graphicx}
\usepackage{textcomp}
\usepackage{xcolor}
\usepackage{subcaption}
\usepackage{biblatex}
\usepackage{algorithm}
\usepackage{algpseudocode}
\usepackage[colorlinks=true,urlcolor=teal,linkcolor=teal,citecolor=teal]{hyperref}

\author{
  Simon Eisenmann\\
  \textit{Technical University of Munich}\\
  Munich, Germany\\
  \texttt{simon.eisenmann@tum.de}
  \and
  Daniel Hein\\
  \textit{Siemens AG, Technology}\\
  Munich, Germany\\
  \texttt{hein.daniel@siemens.com}
  \and
  Steffen Udluft\\
  \textit{Siemens AG, Technology}\\
  Munich, Germany\\
  \texttt{steffen.udluft@siemens.com}
  \and
  Thomas A. Runkler\\
  \textit{Siemens AG, Technology}\\
  \textit{Technical University of Munich}\\
  Munich, Germany\\
  \texttt{thomas.runkler@siemens.com}
}

\title{Model-based Offline Quantum Reinforcement Learning}

\begin{document}

\maketitle

\begin{abstract}
This paper presents the first algorithm for model-based offline quantum reinforcement learning and demonstrates its functionality on the cart-pole benchmark. 
The model and the policy to be optimized are each implemented as variational quantum circuits.
The model is trained by gradient descent to fit a pre-recorded data set. 
The policy is optimized with a gradient-free optimization scheme using the return estimate given by the model as the fitness function.
This model-based approach allows, in principle, full realization on a quantum computer during the optimization phase and gives hope that a quantum advantage can be achieved as soon as sufficiently powerful quantum computers are available.
\end{abstract}

\section{Introduction}
The way in which the particular strengths of quantum computing can be used for machine learning is subject to current research.
One major area of machine learning is reinforcement learning (RL) \cite{sutton2018reinforcement}. 
This technique is of great practical importance, as it allows learning optimal behavior based on observational data for optimized control policies in robotics and industrial applications \cite{haarnoja2019learning,schaefer2007gasturbine,degrave2022tokamak}. 
Offline RL \cite{lange2012batch,levine2020offline} is particularly suitable for learning policies in real-life applications, as it does not rely on trial-and-error learning like online RL, but uses pre-recorded data to learn. 
While online RL is usually costly because it disturbs the operation of the system for which the control is to be learned, offline RL can often use data already obtained during normal operation or manual tuning so that no extra cost and effort are incurred.

There are already many approaches to quantum reinforcement learning (QRL) \cite{meyer2022survey}, but relatively few for the area of offline QRL. 
We are convinced that offline QRL is of particular importance. 
This is, on the one hand, due to the practical advantage that offline RL offers over online RL. 
On the other hand, the data transfer required in online RL between parts of the algorithms that run on the quantum computer and parts that run on classical computers is currently difficult to fulfill.

A particular challenge in offline RL is to master the extrapolation error, which can cause the learned policy to fail outside the range of the training data \cite{wu2019BRAC,fujimoto2019off,swazinna2021moose}. 
An effective method is to use the data to train a surrogate model of the system to be controlled (the environment) that allows extrapolation with less error.
We believe that purely model-based RL methods that use long roll-outs instead of a Q-function to estimate the return are particularly suitable for this purpose.

In this work, we propose model-based offline QRL.
We use variational quantum circuits (VQCs) with data re-uploading to learn a surrogate model offline from pre-recorded data. 
The model is used to evaluate policy candidates, which are generated and optimized by a standard optimizer.
Using experiments on the classical control task cart-pole balancing, we show this approach can solve the control problem.
We summarize our contributions as follows:
\begin{itemize}
\item  To the best of our knowledge, this is the first work to present {\textsc{Model-based quantum reinforcement learning}}.
\item The experimental results show that the proposed method is successful and can learn policies from offline data, 
\item Which shows that the used VQC can model the environment sufficiently good to allow for model-based policy optimization.
\end{itemize}

\section{Related Work}

\subsection{Online Quantum Reinforcement Learning}

The use of VQCs for RL was introduced by \citeauthor{Chen2020} in 2020 \cite{Chen2020}.
There, VQCs are used to represent the Q-function \cite{sutton2018reinforcement}. 
The algorithm used is DQN \cite{Mnih2015HumanlevelCT} and the resulting QRL algorithm was tested for online RL.
Follow-up studies have expanded on this work by exploring various quantum circuits and algorithms to enhance the efficiency and performance of QRL in online settings \cite{lockwood2020reinforcement,skolik2022quantum,franz2023instabilities}. 
However, these approaches focused primarily on model-free RL, leveraging the immediate feedback from the environment to update the policy.

\subsection{Offline Quantum Reinforcement Learning}

Only recently has offline QRL been considered \cite{Cheng_Zhang_Shen_Tao_2023, periyasamy2024bcqq}. 
This area poses unique challenges due to the absence of direct interaction with the environment during learning. 
\citeauthor{Cheng_Zhang_Shen_Tao_2023} use VQCs to represent the Q-function as previously in \cite{Chen2020}, but use CQL \cite{kumar2020conservative}, a dedicated offline RL algorithm, instead of DQN and test their method in offline RL tasks. 
Similarly, \citeauthor{periyasamy2024bcqq} use VQCs to represent the Q-function and the dedicated offline RL algorithm discrete batch constraint deep Q-learning (BCQ) \cite{periyasamy2024bcqq,fujimoto2019off}. 
Both methods belong to the class of model-free RL and assume discrete actions. 

In classical offline RL, model-free approaches (\emph{e.g.}, BCQ \cite{fujimoto2019off} and CQL \cite{kumar2020conservative}) and model-based approaches are used equally, with the majority of model-based methods using both a model and a Q-function (\emph{e.g.},~MORel \cite{kidambi2020morel}, MOPO \cite{yu2020mopo}), and a minority of purely model-based methods without using a Q-function \cite{swazinna2022comparing}.

\subsection{Our Contribution: A Novel Model-Based Approach}

Our approach can be used for both discrete and continuous actions and uses a VQC as a surrogate model and thus belongs to the class of model-based RL. 
We do not use a Q-function, instead we use the surrogate model to roll out long trajectories and use the cumulative reward on these trajectories as return estimates, as introduced for classical RL in \cite{schaefer2007recurrent} and \cite{deisenroth2011pilco}.

The offline QRL methods presented here are most similar to the classical PSONN \cite{psonnZhang, psonn_hein}, which use classical neural networks for the surrogate model.

\section{The Surrogate Model}
Offline RL is based on the use of a pre-recorded data set to learn a policy without interacting with the environment.
We demonstrate the functionality of our model-based method for offline QRL using the cart-pole balancing benchmark \cite{towers_gymnasium_2023}. 
The state space is four-dimensional, comprising the state variables position $x$, velocity $\dot{x}$, angle $\theta$, and angular velocity $\dot{\theta}$.

\subsection{Offline Data Set}
The data set has been generated on the gym environment \textit{CartPole-v1} from the RL benchmark library \textit{Gymnasium}\footnote{\url{https://gymnasium.farama.org}} and consists of 442 episodes generated by a random policy with a total of 10{\small,}000 observations. 
The episodes terminate after an average of 22.6 steps because the pole falls over and leaves the permitted angular range, \emph{i.e.},~$|\theta| > 0.2095$.
By using the random policy and initializing the cart near the center of the track (\emph{i.e.},~$x \approx 0$) and the pole near the upright position (\emph{i.e.},~$\theta \approx 0$), no data is generated near the boundaries of the track ($|x|=2.4$), particularly not with the pole upright far from the center of the track. 
1{\small,}000 data points each are used for validation and testing, and the remaining 8{\small,}000 data points are used for training.

\subsection{Setup of Variational Quantum Circuits}
The practical realization of VQCs is currently limited to simulation on classical computers due to the nascent stage of quantum computing technology. 
Consequently, our experiments were conducted with small-scale quantum circuits, restricted by the current limitations in the number of qubits. 
We utilize the PennyLane software library \cite{bergholm2022pennylane}, which provides comprehensive tools to build and optimize quantum circuits.
The operational mechanics of a VQC involve encoding input data as rotation angles on qubits, utilizing CNOT gates for entanglement, and applying general rotations to facilitate learnable parameters. 
To enhance the learning capacity of the VQC, we adopted the data re-uploading technique, a method that interleaves data encoding with variational processing steps, thereby enriching the expressiveness and adaptability of the model \cite{Schuld2021}.

\begin{figure}[]
    \centering
    \includegraphics[width=0.75\textwidth]{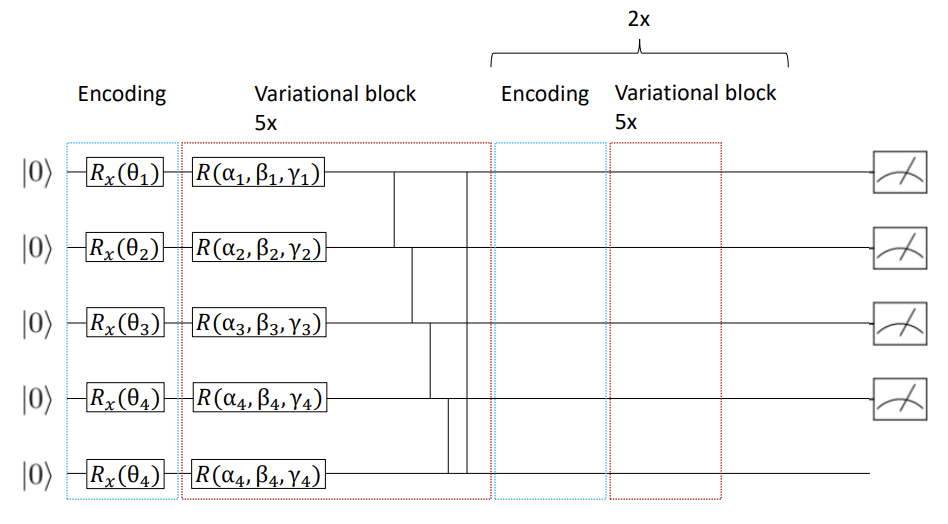}
    \caption{VQC for the cart-pole surrogate model. 
    The design of the sketch is based on the works of \cite{Chen2020} and \cite{bergholm2022pennylane}.}
    \label{fig:vqc_model}
\end{figure}

The model architecture used is a VQC with five qubits, three re-uploadings, and five rotation layers after each data upload (see Fig.~\ref{fig:vqc_model}). 
The four state variables ($x$, $\dot{x}$, $\theta$, $\dot{\theta}$) and the action are used as input. 
All inputs are scaled to the range -1 to 1 and prepared using angle coding.
As targets, the differences between the next state and the current state are used for all four state variables. 
The targets are scaled to the range -0.5 and 0.5.

The training algorithm employed is Adam \cite{kingma2014adam}, utilizing a learning rate of 0.01 and a maximum of 20 epochs. 
The gradients are estimated via the parameter-shift rule \cite{Schuld2018Evaluating, Wierichs2021General}.
To prevent excessively long training times, we refrained from using additional epochs, although increasing the number of epochs might yield better results. 
The parameters from the epoch that show the lowest error on the validation data set are selected.

\section{Model-based Return Estimation}
In our purely model-based approach, we estimate the return (cumulative future rewards) from a given state under a specific policy through simulated roll-outs of model and policy. 
Unlike methods that rely on value or Q-functions, this strategy directly uses the model to predict returns. 
The procedure is as follows: For a given state $s_0$ from a set of start states $S$, the policy $\pi_\omega$ with parameters $\omega$, for which the return ${\cal R}^{\pi_\omega}$ is to be determined, is evaluated and the resulting action $a_0=\pi_\omega({s_0})$ is passed together with the state to the surrogate model $m$, which in turn determines the next state $\tilde{s}_1=m(s_0,a_0)$. 
Note that the tilde symbol is used to emphasize that the model does not necessarily predict the exact next state. 
The reward 
\begin{equation}
    r = \begin{cases} 
    1 , & |x| < 0.5 \land |\theta| < 0.05 \\
    0 , & |x| > 2.4 \lor |\theta| > 0.2095 \\
    0.5 , & \mathrm{otherwise}
    \end{cases}
    \label{eq:reward}
\end{equation}
is calculated from this predicted next state $\tilde{s}_1$ and added to the (undiscounted) reward sum:
\begin{equation}
    {\cal R}^{\pi_\omega}(s_0) = \sum^{H}_{t=1} r(\tilde{s}_t),
\end{equation}
with $s_0\in S$, $\tilde{s}_{t+1}=m(s_t,a_t)$, and $a_t=\pi_\omega(s_t)$.
This procedure is repeated for $H$ steps, where $H$ is the roll-out length, also known as the horizon. 
The reward sum is used as an estimate of the return. 
If the policy or the surrogate model are stochastic, the roll-out should start $N$ times, and the mean reward sum should be used as an estimate of the return.
Here, both the policy and the surrogate model are deterministic, therefore a single roll-out per start state is sufficient.

\section{Searching Optimal Policy Parameters}

The model-based return estimate can be used to optimize policies. 
Since we require the policy to balance from each start state to the end of the episode after 500 steps, we calculate the return estimate over a horizon of 500 steps and average the return estimate over 100 random start states, \emph{i.e.},~$\left| S \right| = 100$. 
We call this value the fitness of policy $\pi_\omega$
\begin{equation}
    {\cal F}^{\pi_\omega} = \frac{1}{\left| S \right|}\sum_{s_0\in S} {\cal R}^{\pi_\omega}(s_0).
    \label{eq:fitness}
\end{equation}

\begin{figure}[ht]
    \centering
    \includegraphics[width=0.75\textwidth]{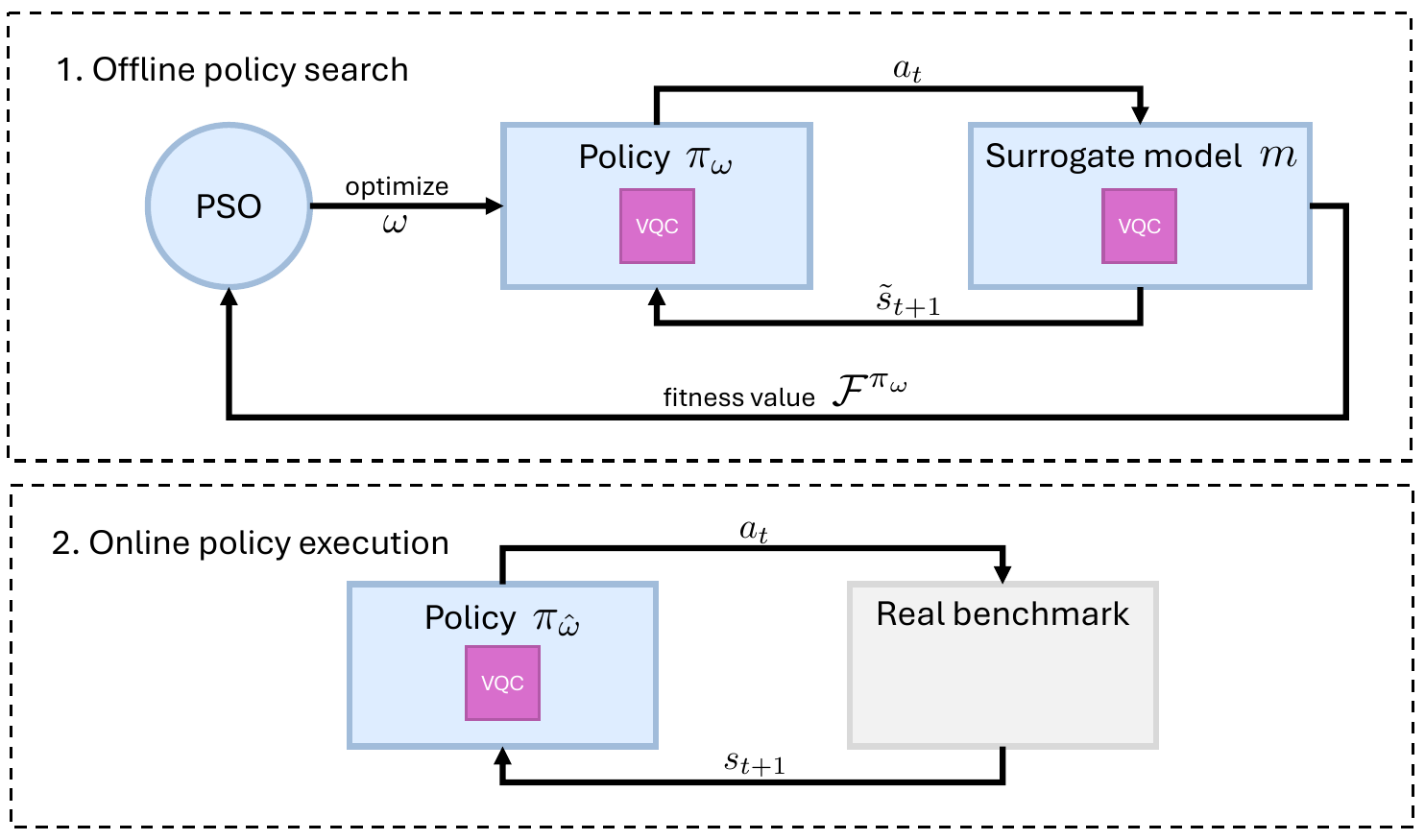}
    \caption{Visualizing direct optimization of policy parameters using PSO in a model-based RL context.}
    \label{fig: policy_search}
\end{figure}

To ensure that the fitness calculation is deterministic, we use the same set $S$ of 100 start states each time. 
The optimization goal is to find a policy that maximizes fitness. 
Because we are aiming for an algorithm that can run entirely on a quantum computer, we also use a VQC as the policy. 
The search for the best policy is thus a search in the space of the parameters of the VQC. 
This could be done in a similar way to classical gradient-based actor-critic methods \cite{sutton2018reinforcement}. 
However, since backpropagation of the gradients is quite slow for VQCs, we use a gradient-free optimization here, which has shown good results in similar settings in classical RL \cite{psonn_hein,swazinna2021WSBC}.

\begin{algorithm}
\caption{\label{algorithm}Quantum Model-Based Reinforcement Learning}
\begin{algorithmic}[1]
\State Initialize environment $Env$
\State Define reward function $r$
\State Initialize surrogate model $m$ with pre-trained weights
\State Initialize particle swarm optimizer $PSO$
\State Start states $S\gets\emptyset$
\State Start state number $N_S$
\For{$i=1$ to $N_S$}
    \State $s_0 \gets Env.reset()$ \Comment{Obtain initial state}
    \State Add $s_0$ to $S$
\EndFor
\State Define optimization function {$\cal F$}({$\omega$})
\Function{$\cal F$}{$\omega$}
    \State Set parameters $\omega$ for policy $\pi$
    \State $totalReward \gets 0$
    \For{$s$ in $S$}
        \For{$t = 1$ to $H$}
            \State $a \gets \pi(s)$
            \State $s \gets m(s, a)$
            \State $r \gets$ $r(s)$
            \State $totalReward \gets totalReward + r$
        \EndFor
    \EndFor
    \State \textbf{return} $totalReward$
\EndFunction
\State $\hat{\omega} \gets PSO.maximize(\cal F)$
\State Test $\hat{\omega}$ using $Env$
\end{algorithmic}
\end{algorithm}

\begin{figure}[]
    \centering
    \includegraphics[width=0.75\textwidth]{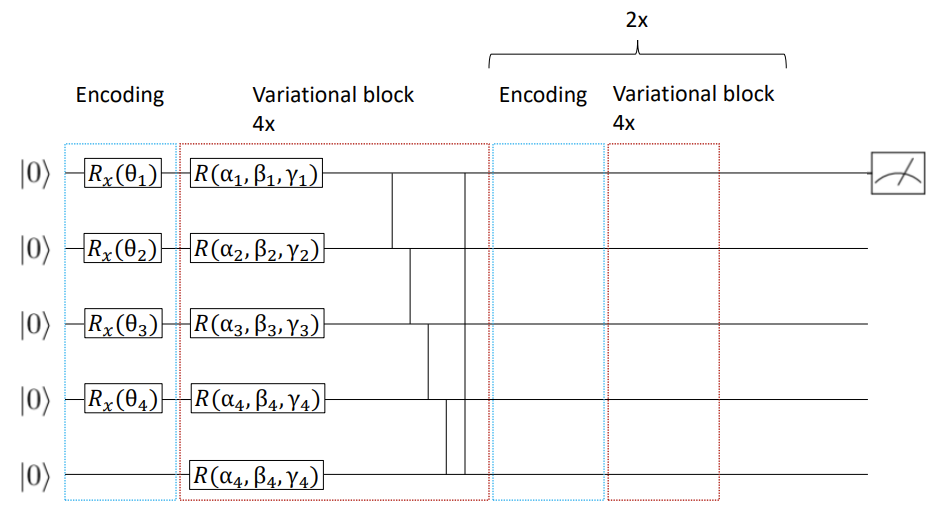}
    \caption{VQC for the policy. 
    The design of the sketch is based on the works of \cite{Chen2020} and \cite{bergholm2022pennylane}.}
    \label{fig:vqc_policy}
\end{figure}

The policy $\pi$ is a VQC with three data re-uploadings, five qubits and four inputs $(x, \dot{x}, \theta, \dot{\theta})=s$, and a total of 180 parameters to be optimized (see Fig.~\ref{fig:vqc_policy}).
To optimize the policy's parameters $\omega$, we use particle swarm optimization (PSO) \cite{PSO} from the Nevergrad library \cite{nevergrad} with 100 particles and a budget of 20{\small,}000 fitness function evaluations. 
Other possible gradient-free optimizers are the Nelder-Mead method \cite{nelder1965simplex}, random search \cite{anderson1953recent}, or simulated annealing \cite{pincus1970monte}.
The policy search algorithm optimizes the policy directly by maximizing the fitness function for the optimal set of parameters:
\begin{equation}
    \hat{\omega}\in\argmax_\omega {\cal F}^{\pi_\omega}.
\end{equation}
The algorithm is visualized in the upper part of Fig.~\ref{fig: policy_search} and summarized in Algorithm \ref{algorithm}.

Note that since PSO is a heuristic algorithm optimizing over a continuous parameter space with a limited optimization budget, there is no guarantee to converge to the optimal set of parameters. 
However, this is an issue inherent to almost every RL algorithm in continuous state spaces.

\begin{figure}[]
    \begin{subfigure}{0.32\textwidth}
        \includegraphics[width=1\textwidth]{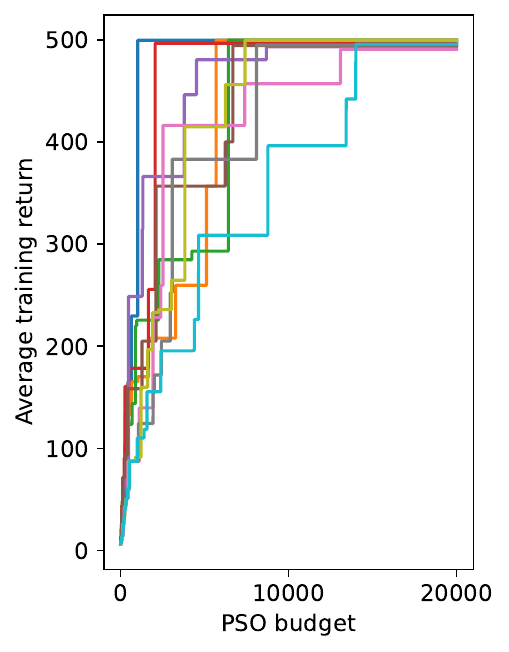}
        \subcaption{Average return on model}
        \label{fig:policy_search_evaluation_a}
    \end{subfigure}
    \begin{subfigure}{0.32\textwidth}
        \includegraphics[width=1\textwidth]{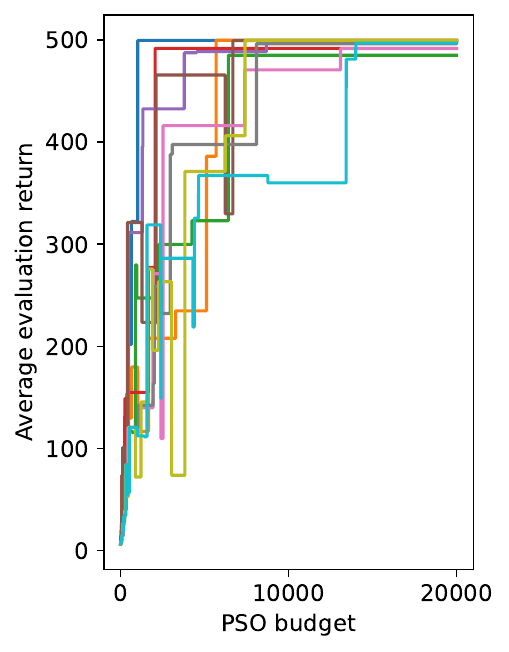}
        \subcaption{Average return on simulation}
        \label{fig:policy_search_evaluation_b}
    \end{subfigure}
    \begin{subfigure}{0.32\textwidth}
        \includegraphics[width=1\textwidth]{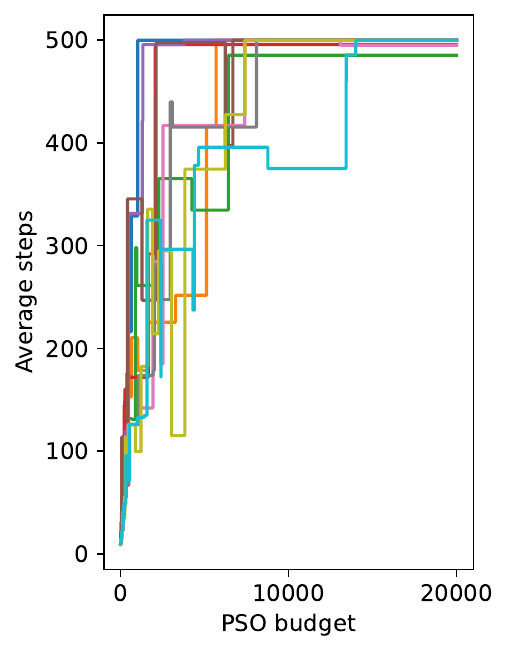}
        \subcaption{Average number of steps on simulation}
        \label{fig:policy_search_evaluation_c}
    \end{subfigure}
    \caption{Learning curves of ten VQC policy search experiments. 
    Same colors equal same experiments. 
    (a) Average training return of the VQC policy search on the VQC surrogate model. 
    (b) Average evaluation return of the same policies evaluated in the cart-pole simulation. 
    (c) Average number of environment steps of the same policies generated in the cart-pole simulation. 
    Note that seven out of ten policies are considered \textit{perfect} policies, \emph{i.e.},~they balanced all of the 100 randomly drawn evaluation states for the full episode of 500 steps.}
    \label{fig:policy_search_evaluation}
\end{figure}

\section{Evaluation}
The goal of model-based policy search for the cart-pole benchmark is to find policy parameters that yield a policy which is able to balance the pole on the gym environment for at least 500 consecutive steps on every possible start state, \emph{i.e.},~states generated by the environment using the reset function.

To test for this capability, we evaluated every improved parameter set during the policy search on 100 start states, which have not been included in the start state set $S$ of the fitness function ${\cal F}^{\pi_\omega}$ (Eq.~\ref{eq:fitness}).

Fig.~\ref{fig:policy_search_evaluation} shows the results of the evaluation of ten individual policy training runs.
Subfig.~\ref{fig:policy_search_evaluation_a} depicts the average training return based on the reward definition of Eq.~\ref{eq:reward}.
The graphs are strictly increasing, since only improved policy parameters during training are considered.
Subfig.~\ref{fig:policy_search_evaluation_b} shows the respective performance of the policy parameters discovered during the training evaluated in the gym environment.
Note that the evaluation return can differ from the training return due to model inaccuracies.
However, we can observe that the VQC surrogate model provides sufficient accuracy to produce well-performing policy parameters after a PSO budget of 20{\small,}000 in all of the ten experiments.

Typically, the performance of cart-pole is reported in average consecutive steps on a large number of randomly drawn initial states.
In Subfig.~\ref{fig:policy_search_evaluation_c}, the average steps on 100 episodes evaluated on the gym environment are depicted.
Note that we consider a policy which balances all of the 100 randomly drawn initial states for at least 500 consecutive steps a \textit{perfect} policy.
Seven out of ten trainings yielded such perfect policies.

From this it can be concluded that
\begin{itemize}
    \item The VQC surrogate model fits the "reality" sufficiently good,
    \item The chosen representation of the VQC policy is sufficient, and
    \item The presented model-based policy search algorithm is functional.
\end{itemize}

\section{Studies on the Surrogate Model}

Since the use of VQCs as a surrogate model is novel, the purpose of this section is to provide more insight into these models. 
We investigate how the prediction quality of the surrogate model depends on the number of data re-uploadings and the amount of data.

\subsection{Effect of Data Re-uploading}

Data encoding can have a major influence on the expressive capacity of the VQC, as demonstrated by \cite{Schuld2021}.
In particular, data re-uploading plays an important role.
In this study, we tested different numbers of uploadings. 
For each number of uploadings, the remaining hyperparameters were optimized with the same computational budget, and the training was repeated 100 times with the respective best hyperparameters. 
The results are shown in Fig.~\ref{fig: Different numbers of uploading 2}. 
The difference between a VQC architecture without data re-uploading and with a single re-uploading is particularly pronounced. 
A single data re-uploading improves the result by more than an order of magnitude, \emph{i.e.},~$13.1\pm0.8$. 
Additional re-uploadings further improve the prediction quality significantly, until the difference between three and four re-uploadings becomes smaller. 
We used a surrogate model with three re-uploadings.

\subsection{Comparing Data Efficiency with Classical Neural Networks}

We investigated how the prediction quality measured in the test set degrades, when the amount of data for training is reduced, and compared it with classical neural networks in each case. 
The classical neural network consists of two layers with 16 hidden neurons, 5 inputs, 4 outputs and the ReLU activation function.
For each of the investigated data sets, 100 trainings were performed for the VQC and the classical neural network. 

\begin{figure}[]
    \begin{subfigure}{0.48\textwidth}
        \includegraphics[width=1\textwidth]{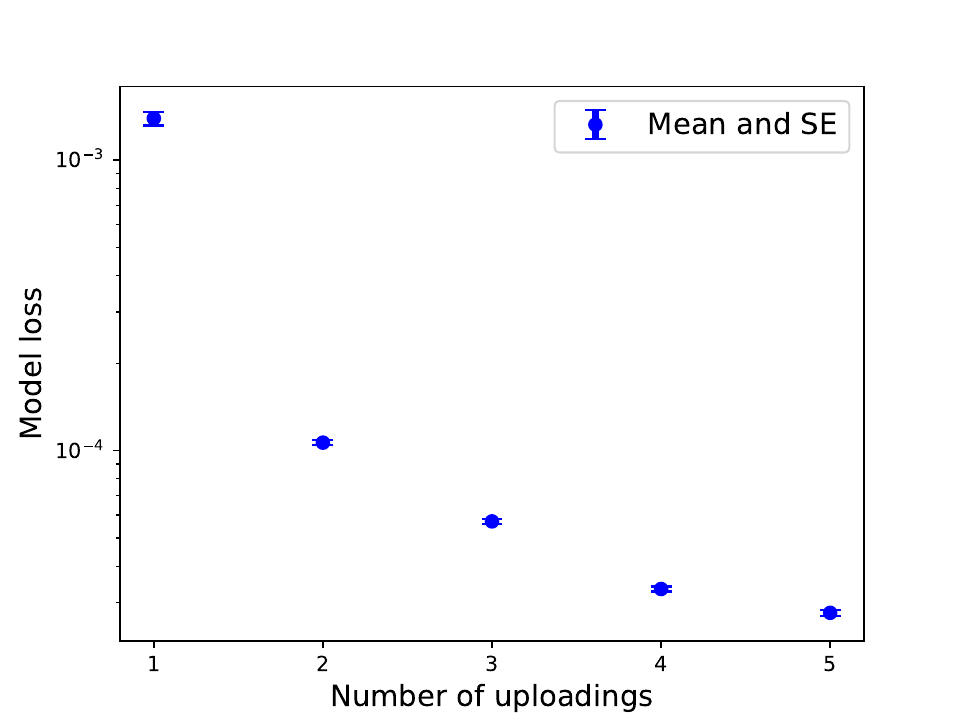}
        \subcaption{Effect of re-uploading}
        \label{fig: Different numbers of uploading 2}
    \end{subfigure}
    \begin{subfigure}{0.48\textwidth}
        \includegraphics[width=1\textwidth]{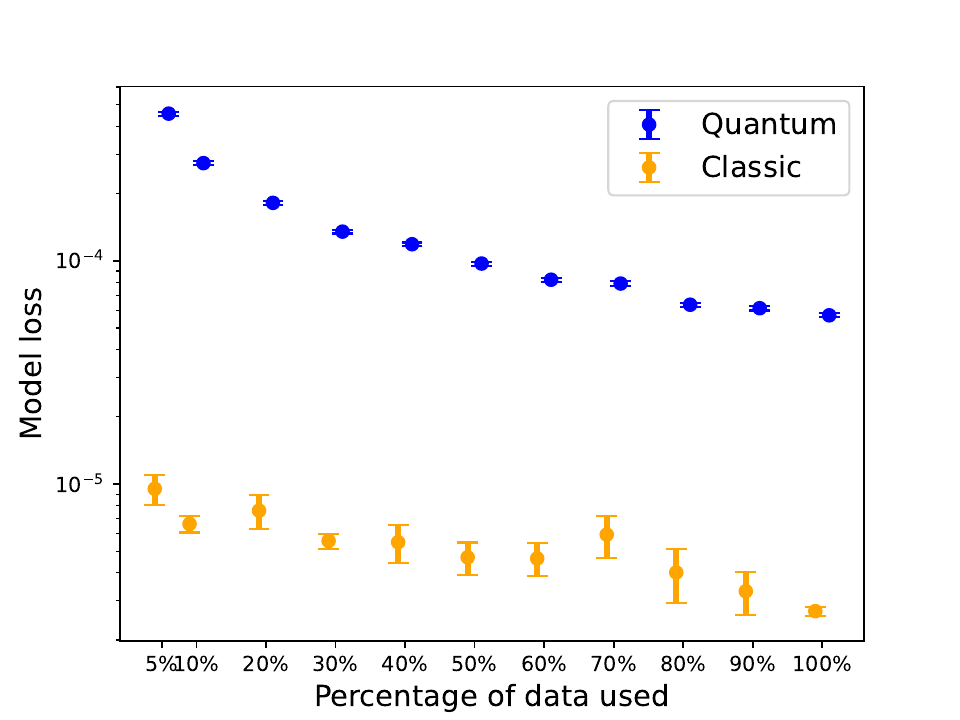}
        \subcaption{Effect of smaller data sets}
        \label{fig: Different numbers of data}
    \end{subfigure}
    \caption{VQC surrogate model experiments. 
    The validation loss is presented. 
    Each point depicts the average over 100 trainings, with $1\sigma$ error bars. 
    (a) Impact of data re-uploading on the prediction accuracy of cart-pole states using a VQC. 
    (b) Comparative analysis of data efficiency between VQCs and classical neural networks in predicting cart-pole states.}
    \label{fig:vqc_experiments}
\end{figure}

It can be seen that, when the amount of data is reduced to one twentieth, the prediction quality of the VQCs becomes $8.0\pm0.3$ times worse, while the prediction quality of the classical neural networks becomes only $3.5\pm0.6$ times worse (see Fig.~\ref{fig: Different numbers of data}). 
Thus, our measurements are in line with the findings reported in \cite{Cheng_Zhang_Shen_Tao_2023} and do not confirm the suggestion of a higher data efficiency of VQCs compared to classical neural networks \cite{Caro_2022,hsiao2022unentangled}.

It is also evident that the results of classical NNs are always far better, being at least $13\pm3$ times better for all data sets. 
Although it is clear that (at least for the cart-pole benchmark) the development of VQCs still has quite a lot to catch up on before it can compete with classical NNs in terms of prediction quality, our experiments (see Fig.~\ref{fig:policy_search_evaluation}) have nevertheless shown that VQCs already model the cart-pole environment sufficiently good to find policies that can reliably master the cart-pole problem.

\section{Outlook - Closed-form Model-based Policy Search}

\begin{figure}[]
    \centering
    \includegraphics[width=0.75\textwidth]{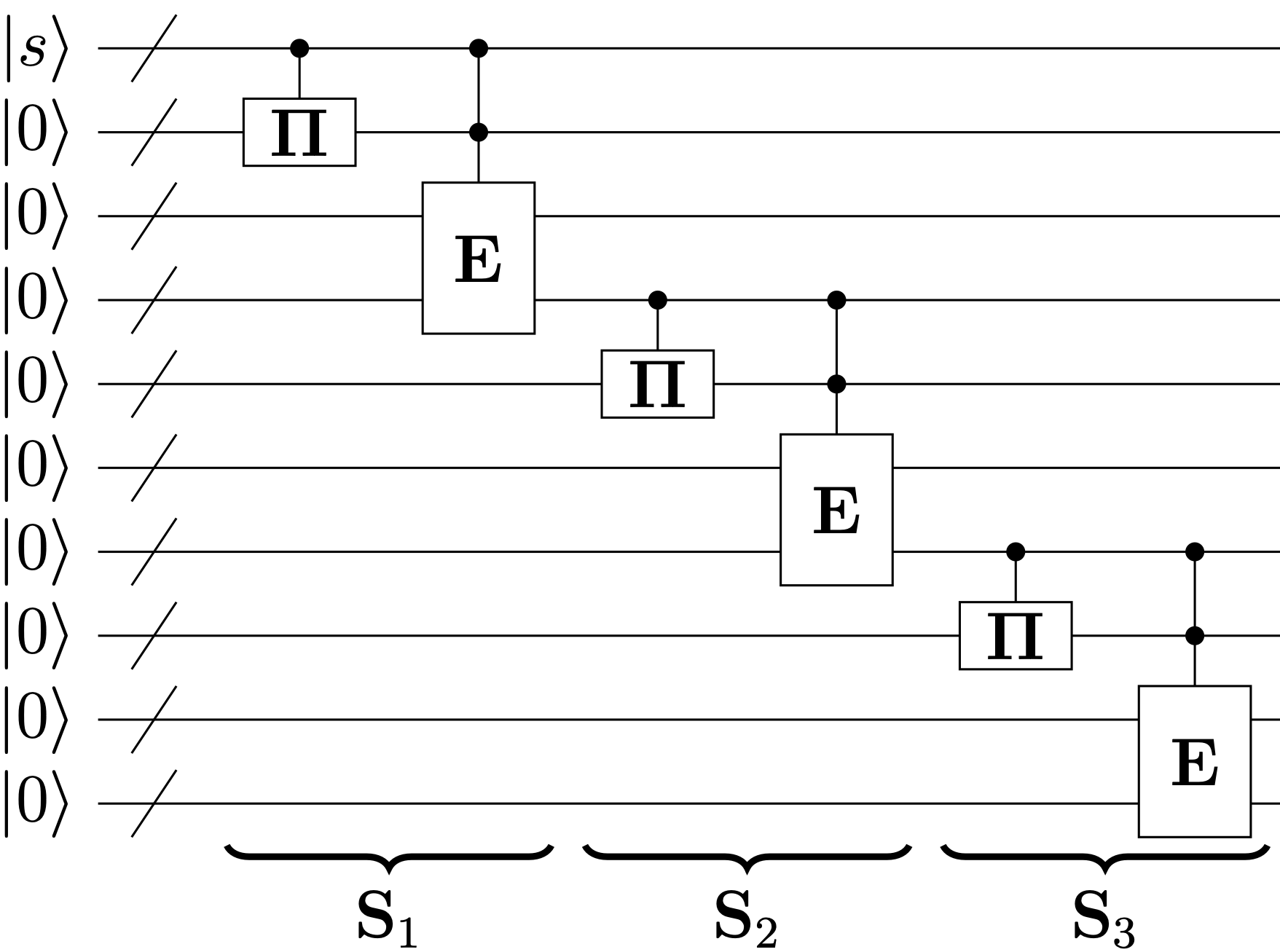}
    \caption{A quantum circuit corresponding to an MDP for horizon $H=3$ as proposed in \cite{wiedemann2023quantum}, showing how a quantum agent ($\Pi$) and a quantum environment ($E$) can be combined to a quantum oracle and utilized for quantum policy improvement using Grover adaptive search.}
    \label{fig:qcrl}
\end{figure}

As both the model and the policy are implemented as VQCs, quantum-based optimization could also be pursued in the future instead of classical optimization (see Fig.~\ref{fig:qcrl}), which could offer a quantum advantage similar to that shown in \cite{wiedemann2023quantum}. 
However, since the entire roll-out would have to be implemented as a quantum circuit, it is currently not possible to simulate such an approach nor to implement it on a real quantum computer.

\section{Conclusion}

In this work, we presented the first model-based approach for QRL and applied it to the cart-pole environment. 
The approach is based on a VQC that is used to learn the dynamics of the cart-pole environment from observational data. 
The VQC learned in this way then serves as a surrogate model. 

The quality of the policy to be optimized is determined by executing the policy for a sequence of steps on the surrogate model, resulting in a roll-out.
A return estimate is calculated from the rewards along this roll-out.
We set the optimization goal to maximize the mean return over a representative set of start states.

We used a second VQC as policy and searched for the best policy in this VQC's parameter space.
For this optimization problem, we used the PSO algorithm, which besides its very good ability to find good optima also works without error backpropagation through the VQCs. 

We demonstrated that our approach is able to find policies that can reliably balance the cart-pole system.
Since the surrogate model is trained using pre-recorded observational data and no interaction with the environment is necessary during optimization, this is also the first model-based {\em offline} QRL method. 

\printbibliography

\end{document}